# Modelling Strategies for the Covalent Functionalization of 2D Phosphorene

Andrea Ienco,* Gabriele Manca,* Maurizio Peruzzini and Carlo Mealli

*Istituto di Chimica dei Composti Organometallici – Consiglio Nazionale delle Ricerche (CNR-ICCOM), Via Madonna del Piano 10, 50019, Sesto Fiorentino (FI), Italy.*

**Abstract**

The paper is an excursus on potential acid-base adducts formed by an unsaturated main group or transition metal species and P atoms of phosphorene ($P_n$), which derives from the black phosphorus exfoliation. Various possibilities of attaining a realistic covalent functionalization of the 2D material have been examined *via* DFT solid state calculations. The distribution of neighbor P atoms at one side of the sheet and the reciprocal directionalities of their lone pairs must be clearly understood to foreshadow the best possible acceptor reactants. Amongst the latter, the main group $BH_3$ or $I_2$ species have been examined for their intrinsic acidity, which favors the periodic mono-hapto anchoring at $P_n$ atoms. The corresponding adducts are systematically compared with other molecular P donors from a phosphine to white phosphorus, $P_4$. Significant variations emerge from the comparison of the band gaps in the adducts and the naked phosphorene with a possible electronic interpretation being offered. Then, the $P_n$ covalent functionalization has been analyzed in relation to unsaturated metal fragments, which, by carrying one, two or three vacant σ hybrids, may interact with a different number of adjacent P atoms. For the modelling, the concept of *isolobal analogy* is important for predicting the possible sets of external coligands at the metal, which may allow the anchoring at phosphorene with a variety of hapticities. Structural, electronic, spectroscopic and energy parameters underline the most relevant pros and cons of some new products at the 2D framework, which have never been experimentally characterized but appear to be reasonably stable.

**Introduction**

Until few years ago, graphene was unique in the field of 2D materials for the electronic properties stemming from the extended π-delocalization. The consequent electrical and thermal conductivity, in association with the low density and high flexibility of the sheets, predisposes graphene for a *plethora* of innovative technological applications.[1] Other 2D materials have been recently investigated, examples being transition metal dicalchogenides[2] or elemental species such as silicene, germanene[3] and phosphorene ($P_n$).[4] The lone pair at each atom of the latter species is apparently inconsistent with a major electron delocalization similar to that of graphene, thus justifying experimental band gap up to 2.0 eV.[5] The gap may be tuned by introducing perturbative effects, a first example being the progressive stacking of the sheets in an ideal reconstruction of the parent black phosphorus material, whose band gap is only 0.33 eV.[6] Incidentally, the latter value is attainable by combining only ten $P_n$ sheets.[4] In any case, in view of the lack of any realistic covalent bond between the layers, the assembly is mainly attributed to the dispersion forces, which also overcome the electron repulsions between the sets of P lone pairs at any two facing sheets. The prevailing attraction is substantiated by the external forces which must be applied to exfoliate black phosphorus, examples being those of micro-mechanical nature, laser irradiation,[4,7] or sonication in solution.[8] Importantly, the tunable band gap for a limited number of stacked layers can be the basis of new useful electronic devices,[9] the drawback being the high sensitivity of the material toward air oxygen and moisture.[10] For this, protection strategies are generally needed, such as for instance the encapsulation of the material in nanocomposites,[11] but also the covering of the $P_n$ surface with acidic chemical groups, including transition metal acceptor fragments. In particular, the anchoring of the latter to the surface may also confer relevant catalytic properties to the system, especially if the metal is not yet completely saturated.

In this paper, we will present systematic analyses, first addressing the chemical and structural aspects of the phosphorene $P_n$ surface and, in particular, the distribution of the P lone pairs, which will interact with σ vacant orbitals at the acidic moieties to be added. The relevance of the donor power at P atoms was already amply discussed by us relatively to the $PI_3$ formation from the $P_4 + I_2$ reactants.[12] It emerged that with respect to a generic $PX_3$ phosphines, the white phosphorus allotrope is definitely a much weaker donor, hence, at this point it becomes important to establish where the phosphorene's lone pairs are positioned compared to the previous limiting cases.

Figure 1 (a and b) starts illustrating the distribution of the $P_n$ lone pairs on one side of the 2D material from two viewpoints.

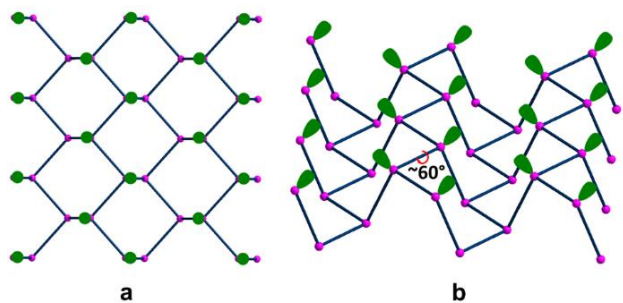

**Figure 1.** Different views of the phosphorene lone pairs at one side of 2D material. The top (**a**) and lateral (**b**) pictures clearly show how the ragged surface consists of ditches in between opposite zig-zag chains with 60° torsions of the two lone pairs associated to a P-P bond.

The top view *a* clearly shows how all the lone pairs point out from a $P_n$ plane although biased by about 60° away from orthogonality, while the lateral view *b* indicates how the two emerging from any P-P bond are reciprocally skewed also by about 60°. Importantly, the rugged $P_n$ surface consists of parallel zig-zag chains with the lone pairs projecting inside the ditches. In principle, any of them may overlap with a mono-functional σ acceptor and, in the case of a metal, contribute to a 2e⁻ donation. The interactions with a multi-functional σ acceptor metal are instead more complex, in view of the simultaneous involvement of two or three neighbor lone pairs, which do not naturally converge into the same point. Still, multiple P-M bonds with one metal seem possible, although they cannot be necessarily equivalent for overlap reasons. Indeed, some geometric adaptation is necessary to allow the multiple-hapticity of phosphorene to a metal.

The mentioned stereochemical aspects will be important in our study of the phosphorene's covalent functionalization, which is still substantially unexplored. In fact, many reports have concerned the van der Waals (vdW) functionalization of the $P_n$ surface, examples being the interaction with either the 7,7,8,8-tetra-cyano-p-quindodimethane (TCNQ) or a perylene diimide,[13] which are characterized by charge transfer from the exfoliated black phosphorus to the organic moiety. Other experimental studies are related to the non-covalent adsorption of small molecules (*e.g.,* CO) on $P_n$, with insufficient stereochemical details about the interactions.[14] One of the few studies of the $P_n$ covalent functionalization with metals has concerned the adduct of the $TiL_4$ tetrahedral fragment (L = sulphonic ester), whose stereochemistry has not been illustrated in detail. In fact, the characterization of the product was performed by liquid NMR-techniques and the presence of a direct P-Ti bond assessed by the Raman spectroscopy.[15] In other cases, the absorption of some naked metal atoms on the surface has been computationally addressed with some focus on the number of coordinated P atoms and the effects on the band structures.[16] Solid state calculations are instead reported to corroborate the absorption of a $CrO_3$ unit on phosphorene with a unique P-Cr bond completing the tetrahedral coordination of the metal. The linkage of 2.45 Å is relatively large although still consistent with covalency, which is further corroborated by the -2.17 eV stabilization energy of the product. No comparison is proposed with any known molecular models of the metal fragment in question.[17]

A detailed experimental and *in-silico* study of the $P_n$ covalent functionalization concerns the micro-mechanically exfoliated black phosphorus over a $Si/SiO_2$ substrate. By reaction with an aryl diazonium salt, which releases $N_2$, formation of P-aryl covalent bonds is described. In our view, there are still open questions concerning the actual electronic nature of the system. For instance, the P-C bonds imply the formation of local phosphonium cations with a predictably associated anion, whose presence has not been addressed. Remarkably, an uncharged system without any counterion must imply unpaired spins, possibly dispersed throughout the Si/SiO$_2$ substrate. In any case, none of the previous points has been focused on at either computational or experimental level.[18]

Based on the summarized status of the art, the covalent functionalization of phosphorene is still bleary, especially from the experimental viewpoint. For this, we propose in this paper a library of potential reactants with proper electronic features to be absorbed at the P$_n$ surface through interactions with selected P lone pairs distributed as in Figure 1. The detailed stereochemical, electronic and energy features of the products appear to be sound for both main group and transition metal reactants, all characterized by residual acceptor capabilities. Initially, our computational approach was based on limited portions of the rugged phosphorene's surface featuring terminal H atoms at the boundaries. However, it became rapidly evident that a model such as P$_{38}$H$_{16}$ could easily undergo geometric distortions that are impossible for an actually periodic P$_n$ surface, hence actual solid state calculations were performed by using CRYSTAL17 package.[19] The increased reliability of the results was verified from structural, spectroscopic and energy viewpoints but also confirmed by the basic electronic properties grasped from the more qualitative molecular modelling.[20] In summary, the two approaches ensured a continuity of interpretation, which did not only shed light on the covalent bonding enabled at the P$_n$ surface but also afforded some predictability, that experimental chemists may be later willing to try.

**Result and Discussion**

**General considerations on the band structure of phosphorene and its affecting parameters**

For a better understanding of the phosphorene's electronic properties and their role in the chemical functionalization, some relevant information may stem from the band structure and the density of states (DOS) shown in Figure 2. The diagrams for the naked 2D material are subdivided in three distinct regions, namely the lower P-P σ bonds, the intermediate frontier P lone pairs and finally the higher and vacant P-P σ* levels. Possible effects on this order are due to structural perturbations[4,6] such as the progressive stacking of P$_n$ sheets in the reconstruction of the bulk black phosphorus. Also, the basic electronic nature will be modified by some acidic molecules or fragments, which may covalently bind with some P lone pairs at the surface. In any case, the original phosphorene band gap changes due to the possible structural deformations as well as the redistribution of the electron density throughout the surface.

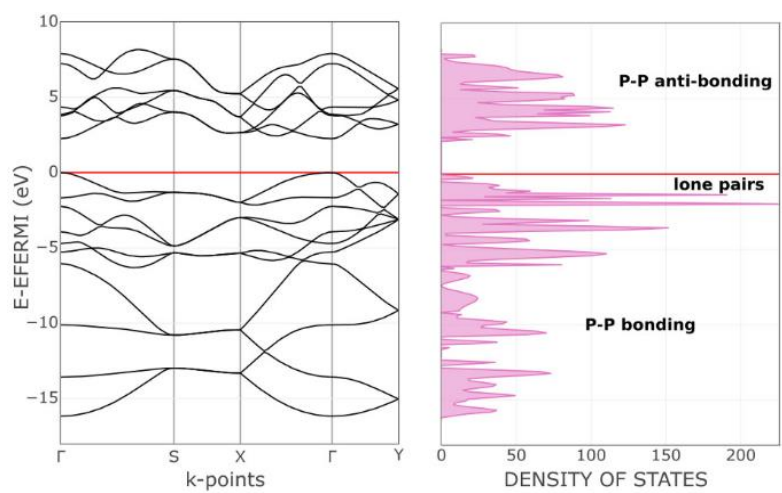

**Figure 2:** Band and DOS structure of one slab of phosphorene in the valence region.

The $P_n$ slab was fixed by assuming a unit cell of four P atoms or sixteen valence orbitals. Accordingly, the overall band structure at the left side of Figure 2 combines sixteen different bands, ten of which are populated by the twenty electrons per cell. In the lowest region, six bands stem from the P-P σ bonds, while the next higher four correspond to the lone pairs of the pyramidal P atoms. Although their orientations exclude convergence, the short contacts across the surface determine reciprocal repulsions, which cause a widening of the frontier band > 1 eV, as estimated by the B3LYP functional.[21] At the same time, a 2.26 eV gap is estimated between the valence and conduction regions, with the latter consisting of six bands having prevailing P-P σ* character. As mentioned, it is sufficient to pile up only ten $P_n$ sheets to approach the band gap proper of the bulk black phosphorus. The steep variation on stacking is likely a consequence of the cumulative bifacial repulsions between lone pairs with consequent widening of the frontier bands.

The covalent functionalization of phosphorene has also electronic and structural consequences on the band structure. For instance, some added reactants may have bands, which fall in between the $P_n$ valence and conduction ones, an example being the case of the d orbital set of an anchored metal atom. In addition, acidic groups at the surface may withdraw electron density from the P lone pairs with consequent decrease of their original repulsion effects. As another aspect, groups bound only at one side of a channel, but projecting over it, may become repulsive toward some opposite P atoms inducing a local physical broadening of the channel itself. Obviously, the more frequent is the repetition of the group along the same zig-zag chain, the more constant is the widening of the channel itself. To evaluate some possible electronic consequences of these structural effects, we started from the optimized structure of the naked 2D phosphorene with the cell parameter *b* (see the inset of Figure 3), which provides an indirect evaluation of the channel width. Then, a systematic stretching of *b* (in the range 4.5 - 4.9 Å) determined the band gap increase from 2.11 to 2.43 eV. A similar effect also occurs upon the stacking of the sheets, as a strategy to reconstruct black phosphorus. This point will be later important for discussing key aspects of the phosphorene's functionalization.

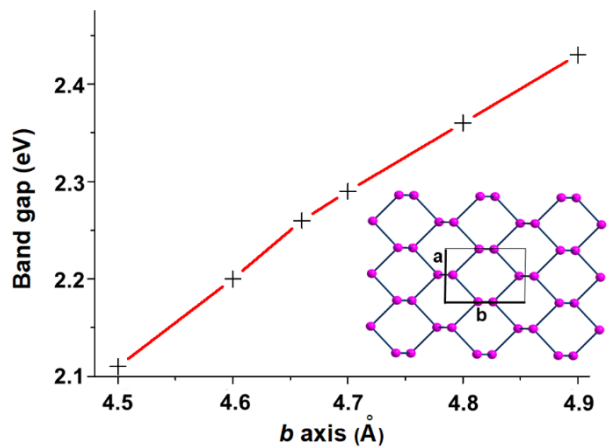

**Figure 3.** The almost linear relationship between the band gap and the cell parameter *b*.

**Non-metal Lewis acids in the phosphorene's functionalization.**

One of the most promising mono-functional acidic molecule for σ interaction with $P_n$ lone pair can be an uncharged borane, also because it excludes the introduction of any charged group at the surface. Boranes are amply reported in the literature to have affinity for phosphorus donors such as phosphanes,[22] and may be also apt for interactions with phosphorene. For our *in-silico* modelling, the $BH_3$ molecule was first chosen to probe its covalent adsorption on $P_n$, hence various solid state adducts of general formula $P_n·x(BH_3)$ were optimized, with x referring to different borane's densities at the surface (see below). In a second stage, other acidic molecules or molecular fragments were taken into consideration, being aware that in some cases an heterolytic splitting of the acidic molecule may induce a local separation of charges. For instance, this could be the case of $I_2$, whose reaction with a strong $R_3P$ phosphine donor leads immediately the ion pair $[R_3PI]^+[I]^-$,[23] whereas the white phosphorus allotrope $P_4$ has very different consequences such as the cleavage of all the six P-P bonds and the formation of the $PI_3$ product after a plethora of intermediates.[12] Also for these reasons, we decided to investigate the behavior of $I_2$ at phosphorene to outline which of the alternative adducts is more likely.

The following discussion will be based on computed parameters, the first ones of geometric characters. Thus, the strength of the donor-acceptor σ interactions in functionalized $P_n$ adducts will be evaluated for instance from the length of the newly formed bonds but also other more or less evident structural variations occurring at either $P_n$ or the added acidic molecule. The other important parameter is the estimated binding energy BE (BE = $\Delta E_{adduct}$ - $\Delta E_{acid-unit}$ - $\Delta E_{P-donor\ unit}$), which helps guessing the propensity of a given molecule to form a stabilizing adduct with phosphorene.

**Phosphorus-borane adducts.**

For comparative purposes, four different types of $P_{pyramidal} \cdot BH_3$ adducts were modelled, the first two involving the molecular phosphines $(CH_3)_3P$ and $CH(CH_2PCH_3)_3P$. The latter with acronym $P_3P$,[24] shown in Figure 4, has a skeleton comparable with that of a local phosphorene subunit, hence with four adjacent lone pairs. Although none of them directly points into another one, repulsion must be at work perhaps even more effective than in the $P_n$ species, where the interactions are spread throughout the entire 2D material.

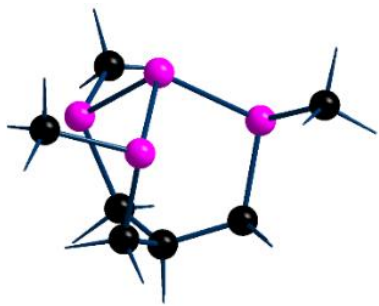

**Figure 4.** Structure of tris(di-*t*-butylphosphino)phosphane, $P_3P$.[24] The methyl group of $^tBu$ substituents are omitted for clarity.

The difference will be corroborated by the comparison of the $P_3P \cdot BH_3$ adduct with the corresponding $P_n \cdot BH_3$ ones, optimized for different borane's coverage at a given side of the surface (see Table 1 and the associated discussion). Finally, a correlation is made with the white phosphorus adduct $P_4 \cdot BH_3$, where, at variance with $P_3P$, there are three additional P-P bonds between the basal atoms of the molecule. In this case, the six edges of the $P_4$ tetrahedron form a highly compact σ bonding system, to which also the four axial P lone pairs are contributing, given that lowest lying $P_4$ bonding combination of $a_1$ symmetry largely consists of $p_z$ in-pointing orbitals, hence lies low in energy. As a consequence, the $P_4$ basicity appears particularly low as previously pointed out by us.[12]

In order to envisage some differences between the adducts, Table 1 reports optimized parameters such as the P-B distance, the BE binding energy and the band gap of the adduct. In particular, the $(CH_3)_3P \cdot BH_3$ one appears most stable in view of the shortest P-B distance of 1.91 Å (1.85 Å in some experimental structure)[25] and the most negative Binding Energy (BE = -1.61 eV). $P_3P \cdot BH_3$ is instead somewhat less stabilized having a larger P-B distance of 1.96 Å and a less negative BE of -1.15 eV. The different basicity is not attributable to a lower energy of the pivotal P lone pair, given the similar destabilizing effects of either the C or P substituents. Rather, the reduced donor power toward the vacant B orbital is due to the minor contribution of the central P lone pair of $P_3P \cdot (HOMO)$, which has instead an about 20% larger mixing of the adjacent P lone pairs in spite of their somewhat unfavorable orientation.

**Table 1.** Optimized P-B distances (Å), BE binding energies (eV) and band bap for various BH$_3$ adducts of pyramidal P donors. In the case of phosphorene P$_n$ different BH$_3$ surface coverages have been considered.

| Model | P-B dist. | BE | Band Gap |
|---|---|---|---|
| (CH$_3$)$_3$P·BH$_3$ | 1.91 | -1.61 | |
| P$_3$P·BH$_3$ | 1.96 | -1.15 | |
| P$_n$·0.031(BH$_3$) | 2.00 | -0.60 | 2.32 |
| P$_n$·0.062(BH$_3$) | 2.00 | -0.58 | 2.42 |
| P$_n$·0.125(BH$_3$) | 2.02 | -0.51 | 2.61 |
| P$_n$·0.250(BH$_3$) | 1.98/2.11 | -0.46 | 2.70 |
| P$_4$·BH$_3$ | 2.08 | -0.32 | |

Before addressing the behavior of phosphorene as a σ donor, it must be mentioned that the white phosphorus adduct P$_4$·BH$_3$ with P-B and BE values of 2.08 Å and -0.32 eV, respectively, represents the weakest adduct of the series. The reason for the very poor 2e$^-$ donor power of P$_4$ were already outlined in our previous study of its demolition with I$_2$, and indeed the highest barrier in the entire process leading to the final PI$_3$ product, was encountered in the formation of the initial P$_4$·I$_2$ adduct.[12] On the other hand, P$_4$ is known to have residual donor power as suggested by some known η$^1$ metal complexes, a first example being the trigonal bipyramidal species (NP$_3$)Ni(η$^1$-P$_4$) species, reported long ago from our institute.[26]

In the case of phosphorene itself, the P-B bonding strength determined for various P$_n$·xBH$_3$ adducts was systematically examined for different x values that each time doubled the density of borane on the surface. Figure 5 shows a top view of the three progressively more covered species, which correspond to one BH$_3$ molecule for every 16 and 8 and 4 P atoms of P$_n$. Additionally, Table 1 reports the parameters also for the least dense adduct (ratio 1:32), where x corresponds to a coverage of about 3.125% *vs*. the other values of 6.25%, 12.50% and 25.0% in the series. Initially, the doubling of the borane coverage causes very little variations of the structural and energy parameters, because the added BH$_3$ molecules do not significantly interfere with each other. Thus, the P-B distance of 2.0 Å remains unaffected and the BE is reduced by only 0.02 eV. In the subsequent passage from 16 to 8 P atoms for any BH$_3$, the reduction of the P$_n$ donor power is evident. In fact, the P-B distance elongates up to 2.02 Å and BE reduces to -0.51 eV.

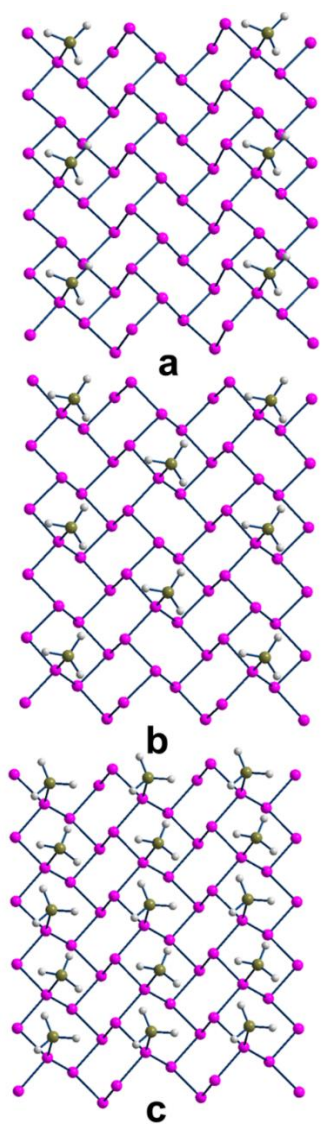

**Figure 5**. The three $P_n$ x($BH_3$) adducts with x=0.062, 0.125 and 0.250 (a, b and c, respectively), corresponding to one $BH_3$ molecule for every 16, 8 and 4 P atoms.

Finally, in reaching the maximum $BH_3$ density (25%), the system is evidently affected, also in view of the two short contacts between neighboring boranes. This determines also an asymmetry of the adjacent P-B distances (1.98 and 2.11 Å) while the difference in BE per $BH_3$ unit becomes as small as -0.46 eV, confirming the increasing repulsions between the absorbates, no more tightly bound.

Figure 6, presenting a lateral view of the adduct with 6.2% $BH_3$ coverage, highlights some $P_n$ deformation. In particular, the channel on which the attached $BH_3$ molecule projects, is more broadened than the adjacent one by about 0.3 Å, thus allowing a larger cradle for the absorbed molecule(s). Recall that a similar broadening of the ditch was simulated for naked phosphorene by elongating the cell parameter *b*, as shown in Figure 3. In that case, the gap between the frontier and conduction bands increased consistently with the results of Table 1 with a maximum 0.5 eV difference

on increasing the BH$_3$ density at the surface. From a qualitative electronic viewpoint, the larger separation of the facing lone pairs at the ditch determines decreased repulsions, the effect being further enhanced by the larger electron drifts into the added boranes. Finally, the narrowing of the lone pair band also implies a larger gap from the conduction band, which, being formed by the P-P σ* levels, is scarcely affected by these geometric perturbations.

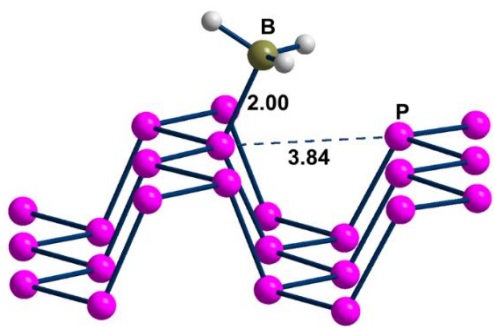

**Figure 6.** Portion of P$_n$·BH$_3$ adduct, where 1 borane molecules is added to a supercell consisting of 16 P atoms.

**Phosphorus-diiodine adducts**

The residual I$_2$ acidity due to its low lying σ* level favors a collinear nucleophilic attack of an external base,[12] such as for instance an axial lone pair of a P$_{pyramidal}$ atom. At variance with the BH$_3$ acceptor, two structural parameters may help evaluating the donor-acceptor interactions with I$_2$, namely a shrinking of the P-I linkage being formed and the corresponding I-I elongation due to the increased population of the I$_2$ antibonding level. The mentioned geometric parameters together with the BE values are reported in Table 2 for the same series of P donors considered in Table 1. Starting with tri-alkyl phosphines, the nucleophilic attack on I$_2$ is most efficient, as proved by the experimental adduct $^i$Pr$_3$P·I$_2$[23] and the optimized (CH$_3$)$_3$P·I$_2$ one.[12] In both cases, the P-I linkage is as short as ~2.4 Å while the I-I distance is largely elongated (3.40 Å *vs.* the 2.73 Å value for the free diatomic). The persisting collinearity of the adducts recalls the classic Halogen Bonding (XB) picture with expected asymmetric linkages.[27] On the other hand, the attacked diatomic can be definitely split with the assistance of a second I$_2$ molecule, which extracts the terminal iodide affording the ion pair [Pr$_3$P$_3$I]$^+$ [I$_3$]$^-$. The corresponding distances in Table 2 are slightly less pronounced than those already determined by us through a different computational approach,[28] while the present BE value of -0.98 eV is consistent with the previously calculated ΔG of -23.5 kcal mol$^{-1}$ (-1.01 eV),[12] which includes the penalizing entropy term. The latter should be almost constant for the whole series in Tables 2. By switching to the phosphine adduct P$_3$P·I$_2$, the diatomic is somewhat less activated given the P-I and I-I distances of 2.86 and 3.03 Å, respectively. The point is also corroborated by the smaller BE value *vs.* (CH$_3$)$_3$P·I$_2$ (-0.65 *vs.* -0.98 eV) implying a minor electron transfer from the P lone pair into I$_2$ σ*.

As stated above, the P$_3$P weaker donor character is due to the about 20% mixing into the HOMO of the three lateral P lone pairs with the central one.

**Table 2**. Optimized P-I and I-I distances (Å) and the BE binding energies (eV) for the selected series of P$_{pyramidal}$·I$_2$ adducts analogous to those in Table 1. For simplicity, the uniquely examined P$_n$ adduct contained one I$_2$ molecule for any 16 P atoms at the surface.

| P$_{donor}$ | P$_{pyram}$·I$_2$ | | |
|---|---|---|---|
| | P-I | I-I | BE |
| (CH$_3$)$_3$P | 2.78 | 3.07 | -0.98 |
| P$_3$P | 2.86 | 3.03 | -0.65 |
| P$_n$ | 3.17 | 2.94 | -0.18 |
| P$_4$ | 3.19 | 2.92 | -0.13 |

The calculations confirm a very weak interaction in P$_4$·I$_2$, as suggested by the almost unperturbed geometries of the two molecular components and the very small -0.13 eV exothermicity. In a previous study of the multi-step demolition of P$_4$ with I$_2$ to provide PI$_3$, the evolution from the first adduct P$_4$·I$_2$ was the most difficult one for the entire process in view of the high and unique barrier ($\Delta$G = +14.6 kcal mol$^{-1}$). In actuality, a second I$_2$ molecule had to be involved as well as in all the subsequent steps.[12] In this manner, two distinct P-I linkages could be formed in place of the pre-existing P-P bond, while a new I$_2$ molecule was regenerated *in situ* being ready for subsequent reactivity. With this picture in mind, we explored whether also phosphorene could undergo any similar P-P cleavage under the action of two diatomics. Again we started with the XB type adduct P$_n$·I$_2$ where one diatomic interacts with a P$_n$ cell of 16 P atoms with an activation that is only marginally larger than in P$_4$·I$_2$ (differences in the P-I and I-I distances of about 0.2 Å and a slightly more exothermic energy balance of -0.18 *vs*. -0.13 eV). Upon the addition of a second I$_2$ molecule, the optimized P$_n$·2I$_2$ model, reported in Figure S1, shows that the di-iodine activation only barely increases. In fact, the first added I$_2$ molecule has P-I$_1$ and I$_1$-I$_2$ variations no larger than 0.05 Å and the same result applies also to the second I$_2$ one (see Table S1). In this case, the overall BE value indicates an extra energy gain of only -0.12 eV, but more important a further evolution of the system seems prevented by an unsuitable stereochemistry. While in the P$_4$ case, it was evident that remote I atom of the 2I$_2$ grouping can potentially perform a nucleophilic attack into P-P $\sigma$* level and induce its cleavage,[12] the same does not apply to the 2D P$_n$ species in view of the unsuitably oriented P-P bonds. Perhaps, the mechanism could be pursued for the high temperature transformation of the red phosphorus allotrope into the black one thanks to the involvement of $I_2$ molecules generated by $SnI_4$.[29] In this case, the stereochemistry of the P-P linkages of red phosphorus is not as flat as in phosphorene possibly favoring the action of $I_2$ similarly to that proposed for $P_4$. Obviously, the problem has to be tackled in some depth in order to corroborate such a conclusion.

**Transition metal fragments for covalent anchoring at phosphorene.**

Phosphorus based ligands are ubiquitous in organometallic chemistry and have potential relevance in catalytic processes. In principle, also one or more adjacent pyramidal P donors of phosphorene could coordinate an unsaturated metal center associated to a number of external coligands. This implies the covalent functionalization of the 2D material with metal centers, a subject scarcely tackled up to now at both the experimental and theoretical levels.[15,16,17,18] The new $P_n$-M bonding may in principle ensure full metal saturation, although a residual unsaturation may become relevant to support a catalytic behavior of the species anchored at the surface.

To choose the metal fragments, which may be best suited for the phosphorene functionalization, it is first important to have a good understanding of the stereochemical and electronic properties of both the interacting phosphorene and transition metal fragment. In the choice of the latter, a leading concept is that of the *isolobal analogy,*[30] which allows their tailoring to support single or multiple $2e^-$ interactions with neighbor $P_n$ lone pairs. A general strategy is that of starting from a formally saturated metal complex, which, on losing one or more coligands, acquires a variable number of vacant σ hybrids acting as the acceptors of some phosphorene P lone pairs. The donor power of the latter has already been found to be particularly weak, hence it is important that the original metal complex carries some even weaker ligands to be eventually replaced. As another problem, the generated metal fragments should not create peculiar steric problems in the approach to the 2D material but also in their reorientation on the surface to maximize the σ overlap(s). Another important limitation in the choice of the metal fragment is its possible charge derived from the combination of different metal and coligands. In this case, the necessary electroneutrality of the solid state system imposes the presence of a counterion in close proximity of the charge metal fragment anchored at the surface. Clearly, for an optimal metal selection in the $P_n$ functionalization, one should preferentially work with an uncharged metal fragment.

Figure 7 shows top views of one phosphorene's face with differently hanging metal fragments. The latter differ for the number of σ hybrids stemming from the metal and interacting with one, two or three $P_n$ lone pair, as shown in Figure **7a-c**. Since any P lone pair forms an angle of about ~30° with the $P_n$ sheet, the unique $P_1$-M linkage of the $\eta^1$ coordination (**7a**) is expected to be bent on the surface, although in some cases some reorientation is observed (see below).

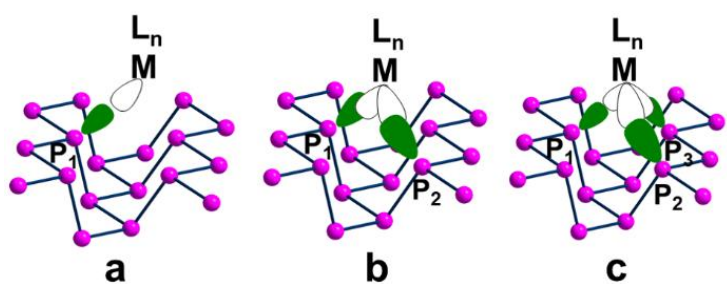

**Figure 7**. A phosphorene channel with differently anchored $L_nM$ fragments: **a)** $\eta^1$ coordination of $P_n$ with a single σ acceptor metal; **b)** $\eta^2$ coordination at a doubly (*cis*) unsaturated metal fragment; **c)** $\eta^3$ coordination at a triply (*fac*) unsaturated metal fragment.

In the case of the $\eta^2$ and $\eta^3$ coordinations (**7b** and **7c**, respectively) not all the interactions imply an optimal orbital overlap. In the former case, for instance, the lone pairs at the $P_1$ and $P_2$ atoms, originally lying in parallel planes at the opposite sides of a ditch (see Figure 1), need to interact simultaneously with a midway M atom. To maximize the overlap some local torsion should occur at the surface, which has however a very scarce degree of flexibility. A similar problem arises for the $\eta^3$ coordination, which involves one atom at one bank of the ditch ($P_1$), while the $P_2$ and $P_3$ one are sequential at the opposite side. In this case, only the $P_1$ lone pair can in principle attain an optimal overlap with one M σ vacant hybrid, conversely, those at the $P_2$ and $P_3$ one should attempt a rather difficult reorientation. The problem will be later confirmed by the larger $P_2$-M and $P_3$-M bonding distances compared to the $P_1$-M one.

## $\eta^1$ Coordination Mode

Possible examples of the general transition metal fragments featuring a singly vacant σ hybrid are shown in Scheme 1. Tentatively, we selected some general models to test in each case the possible $P_n$ functionalization. Once again, we defined a cell of sixteen P atoms for each added metal fragment to test *in-silico* the most relevant aspects of this chemistry.

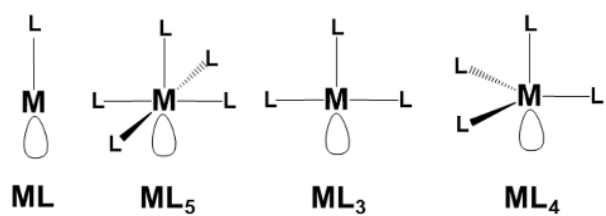

**Scheme 1.** Generalized metal fragments in principle suited for the $\eta^1$ coordination of phosphorene.

**$d^{10}$-ML linear fragment.** The $\eta^1$-$P_n$ covalent functionalization was best verified by using the $d^{10}$-AuCl fragment. In this case, no major steric hindrance problem seems to be at work in the linear alignment of one $P_n$ lone pair with the Au(I) vacant σ hybrid. The uncharged Au(I)Cl fragment can be in principle derived from a linear complex such as $AuCl_2^-$ [31] or ClAuL upon the extraction of a Cl⁻ anion or a L neutral ligand such as $Me_2S$[32] or $R_3P$.[33] Given the about 30° bias of any P lone pair on the average $P_n$ plane, the $P_4$-$P_1$-Au angle of 114° of the optimized ClAu($η^1$-$P_n$) adduct (Figure 8) is close to the ideal 110° value, thus ensuring an almost optimal overlap between the metal and P σ orbitals. The point is corroborated by the 2.24 Å distance of $P_1$-Au bond, which is smaller than the sum of the covalent radii (2.43 Å)[34] and close to the average single bond derived from Cambridge Database structures.[35] Any major steric hindrance problem seems excluded by the large separation of about 3.85 Å between the terminal chloride ligand and the $P_2$ and $P_3$ atoms at the opposite bank of the channel. While the $P_2$-Au and $P_3$-Au distances are as large as 3.04 Å, it cannot be excluded that a residual attraction is at work.[36] An indication may be the about 2° deviation from linearity of the $P_1$-Au-Cl angle and the consequent biasing of the metal toward the P atoms at the other side of the ditch. The latter is about 0.2 Å widened with respect to free phosphorene, suggesting that the channel adapts for a better accommodation of the projecting fragment, with an effect on the phosphorene band structure. The specificity of the latter in this case will be discussed below.

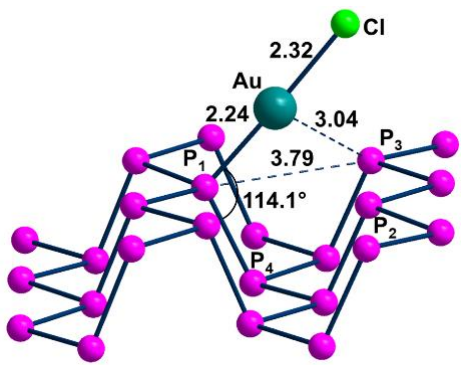

**Figure 8.** The optimized adduct ClAu($η^1$-$P_n$).

The stabilization of the ClAu($η^1$-$P_n$) adduct is proved by the large BE value of -2.4 eV, which is about four times larger than the -0.58 eV value obtained for the non-metal species $P_n$ $BH_3$ in Table 1. In view of these results, it emerges that an opportune transition metal fragment can be one of the best candidates for the covalent functionalization of phosphorene. For an even more realistic energetic picture, the ClAu($η^1$-$P_n$) adduct has been obtained through two alternative paths. Eq. 1 involves the precursor ClAu($Me_2S$), which first loses $Me_2S$ with a cost +2.2 eV, hence the final energy balance is exothermic by -0.2 eV. Remarkably, the adduct formation from the other mentioned complex ClAu($Me_3P$) is more difficult, since Eq. 2 is endothermic by +0.44 eV. Thus, phosphorene functionalization can be generalized to proceed from reactants carrying a good leaving group.

ClAu(Me$_2$S) + P$_n$ → ClAu($\eta^1$–P$_n$) + Me$_2$S          Eq. 1

ClAu(Me$_3$P) + P$_n$ → ClAu($\eta^1$–P$_n$) + Me$_3$P          Eq. 2

It is worth mentioning that previous computational studies on gold-phosphorene were reported for a system with single Au(0) atoms dispersed on the surface and $\eta^3$ locally bound to the cavity formed by the P$_1$, P$_2$, P$_3$ atoms.[16] The average optimized Au-P distances of about 2.37 Å and the -1.61 eV stabilization energy seemed to confirm the propensity of gold toward P$_n$ functionalization but the peculiar electronic nature of the Au(0) carrying an unpaired spin each was not specifically illustrated.

At this point, it is worth mentioning that, based on the *in-silico* predictions about the ClAu($\eta^1$-P$_n$) adduct, some experimental attempts of obtaining a functionalized phosphorene surface was carried out by us starting from the ClAu(Me$_2$S) reactant. Unfortunately, we still unable to present here any unquestionable result, because ClAu(Me$_2$S) easily promotes the formation of gold nanoparticles.[37] During the reaction, we indeed observed by-products, some of which have been spectroscopically characterized. The work, which is still in progress, will hopefully provide some reasonable explanation of the Au(I)→Au(0) reduction. To avoid the latter, we are trying to use some Au(I) linear complex other than ClAu(Me$_2$S) for the P$_n$ functionalization

Concerning the anticipated effect on the P$_n$ band structure induced by the functionalization with AuCl, we noticed a decrease of the band gap, rather than the increase expected for the widening of some ditch. Indeed, the 2.26 eV band gap of free phosphorene becomes 2.17 eV in ClAu($\eta^1$–P$_n$). A relatively easy explanation of the result emerges from the plots of the whole band and the corresponding Density of States (DOS) in Figure 9.

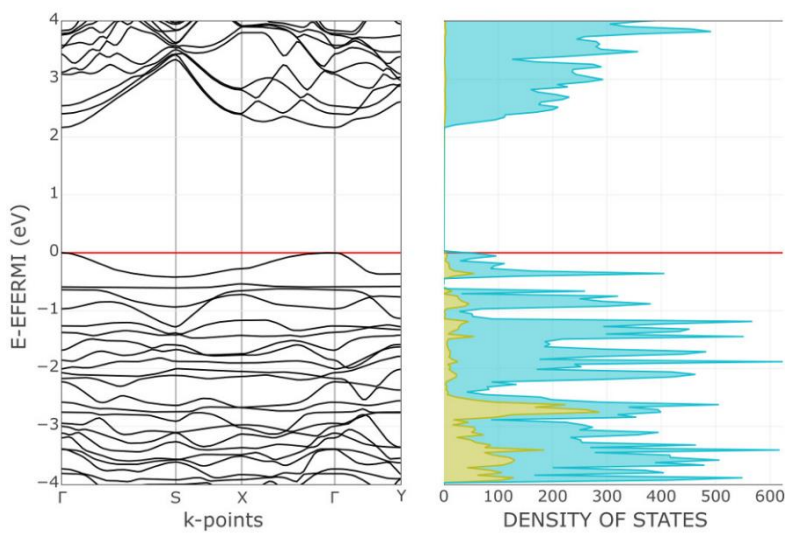

**Figure 9.** Band and DOS structure for the adduct of ClAu fragments to phosphorene (1:16 ratio). Near the Fermi level (red line), a significant contribution from gold d orbitals in yellow falls intermediate between the valence and conducting bands with consequent narrowing of band gap.

Evidently, the highest portion of occupied bands (Fermi region) largely consists of the set of Au d orbitals (yellow projection), which are implicitly repulsive with P lone pairs. Importantly, the spread of the basic band structure determines a consequent reduction of the gap with respect to the conduction band. The point is further corroborated by a plot of the Crystal Overlap Orbital Population (COOP in Figure S2), which properly highlights the triggered d/P lone pairs repulsions.

**$d^6$-ML$_5$ square pyramidal (SP), $d^8$-ML$_3$ T-shaped and $d^8$-ML$_4$ trigonal pyramidal (TP) metal fragments.** The behavior of the additional metal fragments in Scheme 1, all carrying a uniquely vacant σ hybrid to be potentially exploited in the $\eta^1$-P$_n$ coordination, is briefly addressed here. For instance, a $d^6$–ML$_5$ SP fragment derived from a classical octahedral complex of group VI (M(CO)$_6$ with M=Cr, Mo, W)[38] may potentially interact with a P lone pair of the P$_n$ surface upon some bending on the surface, which maximizes the σ overlap. In this case, however, close contacts are formed between P atoms and at least two basal CO ligands of the metal fragment with consequent steric effects. Something analogous occurs for a T-shaped fragment such as PtCl$_2$(CO), which formally descends from the square planar $d^8$ complex PtCl$_2$(CO)$_2$[39] or a Trigonal Pyramidal (TP) one (*e,g.,* $d^8$-(CO)$_4$Ru) derived from a TBP precursor such as (CO)$_5$Ru[40]). In actuality, all the mentioned P$_n$ adducts were optimized, but in their structures, shown of Figure 10a-c, the metal fragment is not bent but almost upright on the surface. This is confirmed by the P$_2$-P$_1$–M angles, which are about 25-35° more open than the ideal 110° value, hence far from allowing maximum σ overlap. Consequently, the P$_1$–M bonds are weaker than expected, as confirmed by the optimized (M= Mo, Pt, Ru) distances, which are in general about 0.1 Å larger than single bonds.[35] Most likely, the adopted geometry helps preventing short contacts between some coligand and P atoms on the surface. On the other hand, all of the structures of Figure 10 are actual minima, which imply a certain degree of stability in spite of their scarce P-M direct bonding. A possible explanation may be based on pointed out short vector(s) between some surface P atoms and the upper CO ligand. Remarkably, the latter is not exactly collinear with the bound metal atom but features a M-C-O angle bent up to about 15°. This can be suggestive of an incipient non-covalent P···C(O) interaction, occurring between a P lone pair and a vacant CO π* level mainly at the side of the C atom. Even weaker interactions of this type are likely at work in the adsorption of gaseous CO on phosphorene, as suggested by other authors.[14] Since points like this are not in line with the presently studied covalent functionalization of phosphorene, the electronic features of the intriguing compounds of Figure 10 have been no further explored.

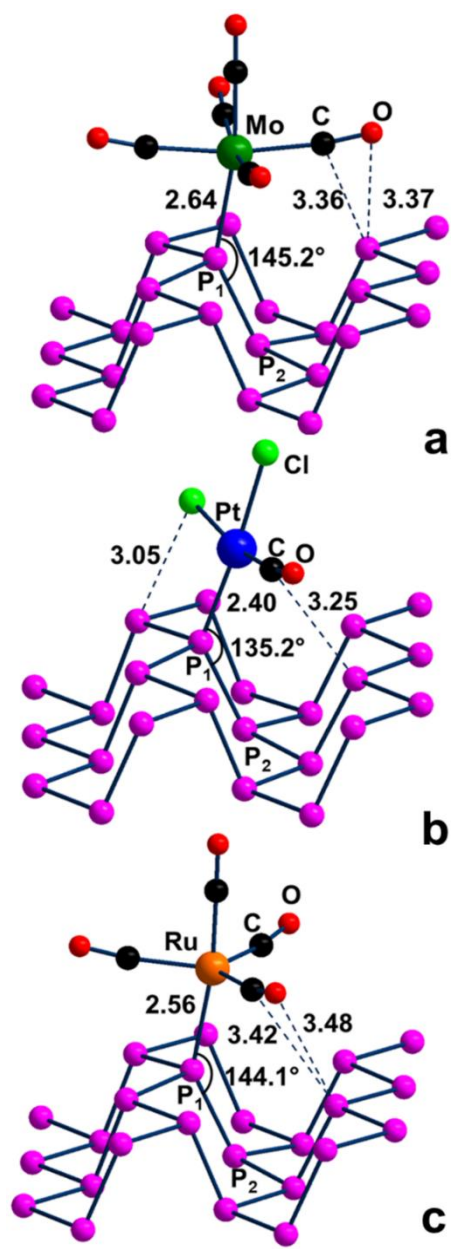

**Figure 10.** Optimized adducts between $P_n$ and different metal fragments with a single σ acceptor function: **a)** $d^6$-(CO)$_5$Mo Square Pyramid, SP; **b)** $d^8$-Cl$_2$(CO)Pt T-shaped fragment; **c)** $d^8$-(CO)$_4$Ru trigonal pyramid (TP).

**Poly-hapto coordination of a metal by phosphorene.**

The phosphorene's sequence of interlinked P atoms, each carrying a single σ lone pair, suggests the possibility of bis-chelate or tris-chelate behaviour of the 2D material. The functionality of phosphorene as a multiple donor must be related to the disposition of the selected close lone pairs at the $P_1$, $P_2$ and $P_3$ atoms, as outlined in Figure 7. Again, the identification of suitable metal fragments able to accept simultaneously two or three 2e$^-$ donations can be gained from the application of the *isolobal analogy* concept.[30]

## η² Coordination Mode

L$_2$M fragments, derived from square planar d$^8$ complexes upon removal of two uncharged *cis* ligands, carry two vacant σ hybrids suitable for the η² coordination of phosphorene across one of the channels. For instance, a suitable precursor can be a Ni(II) complex with two methyl ligands and two uncharged 2e$^-$ donors (*e.g.,* H$_2$O), which on losing the water molecules, allows the orbital interactions of Figure 7b in the product (CH$_3$)$_2$Ni(η²–P$_n$). An optimization of the latter species, by using again the 1:16 Ni:P ratio, confirmed that the P$_1$ and P$_2$ atoms across a P$_n$ channel can also complete the square planar coordination of the metal as in the structure of Figure 11. Here, the optimized Ni-P distances of 2.24 Å are about 0.1 Å longer than in the comparable X-ray structures formed by a diphosphine chelate,[41] while the P$_1$-Ni-P$_2$ angle of 95.2° is about 8° more open. These aspects are likely the consequence of the already known weaker donor power of the phosphorene P atoms and their strained geometry.

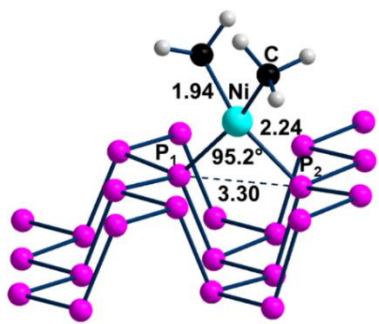

**Figure 11**. Optimized structure of adduct (CH$_3$)$_2$Ni(η²-P$_n$) with Ni:P ratio of 1:16.

The simultaneous coordination of the two P atoms at the opposite sides of a channel imposes a shrinking of the latter, which is indicated by the 0.2 Å shorter P$_1$⋯P$_2$ separation compared to naked phosphorene. Correspondingly, the adjacent channel is somewhat widened. The P$_1$ and P$_2$ lone pairs involved in the η²-P$_n$ coordination lie as known in parallel planes, hence do not naturally converge into the unique metal position with non optimal overlap with its vacant σ hybrids. For this, the metal fragment is rotated by about 25° with respect to the perfect square planar geometry. Scheme 2 helps to understand the orbital underpinning of the distortion, needed to improve the σ overlaps.

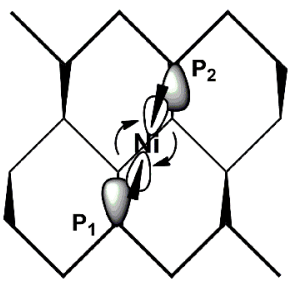

**Scheme 2.** Suggested rearrangement of a L$_2$Ni(II) fragment across one P$_n$ channel to maximize σ overlap.

Through some minor deformation of the phosphorene surface and the indicated metal rearrangement, the obtainment of the final $P_n$ adduct is indeed exotermic by -0.32 eV, as estimated from the reaction in Equation 3.

$$(CH_3)_2Ni(H_2O)_2 + P_n \rightarrow (CH_3)_2Ni(\eta^2\text{-}P_n) + 2\ H_2O \qquad \text{Eq. 3}$$

The *quasi* square planar coordination of $d^8$-Ni(II) on $P_n$ determines an increase of the band gap with respect to that of naked phosphorene (from the 2.26 to 2.31 eV, as shown in Figure S3). The trend is opposite compared to that of the ClAu derivative. There must be a compromise between the reduced repulsion between the $P_1$ and $P_2$ lone pairs participating in metal binding and the larger repulsion between the not involved ones due to the shrinking of the channel.

The $\eta^2$ coordination of phosphorene may be potentially attained also by using alternative fragments. One of them can still be of the $L_2M$ type with a $d^{10}$ rather than a $d^8$ metal atom, which has as a precursor a 18e⁻ tetrahedral complex ($T_d$). In this case, the two vacant σ hybrids of the $L_2M$ fragment lie high in energy for having exclusive s/p character without any d contribution. In general, the known Ni, Pd, Pt tetrahedra are stabilized by strong σ donors such as four COs[42] or 2COs + 2R$_3$P ligands.[43,44] Since some complex is known with a dinitrogen chelate,[45] we used the species $(CO)_2Ni(NH_3)_2$ as model reactant for the reaction with $P_n$, reported in Eq. 4. From the optimized adduct $(CO)_2Ni(\eta^2\text{-}P_n)$, having local tetrahedral geometry as shown in Figure 12, the process is estimated to be exothermic by -0.17 eV.

$$(CO)_2Ni(NH_3)_2 + P_n \rightarrow (CO)_2Ni(\eta^2\text{-}P_n) + 2NH_3 \qquad \text{Eq. 4}$$

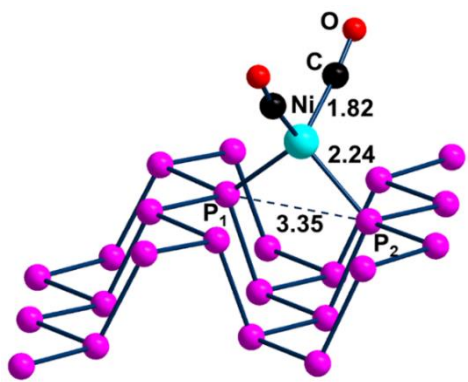

**Figure 12.** Optimized structure of the $(CO)_2Ni(\eta^2\text{-}P_n)$ adduct.

In the tetrahedral geometry of $(CO)_2Ni(\eta^2\text{-}P_n)$, the Ni(0)-P distances of 2.24 Å are close to those of the Ni(II) square planar species of Figure 11. Therefore, the channel width is again shortened by about 0.2 Å but the 2.22 eV band gap of the $d^{10}$ adduct is now only slightly smaller than the 2.26 eV value of free phosphorene *vs.* the 2.31 eV value of the Ni(II) case. Figure S4 shows trends similar to those of the ClAu adduct in view of the position of the metal d orbitals in highest portion of the valence band. A final piece of information concerns the about 70 cm$^{-1}$ blue shift of the CO ligands found in the computed IR spectrum of $(CO)_2Ni(\eta^2-P_n)$ compared to that of the molecular model $(CO)_2Ni(PMe_3)_2$. This feature is not surprising given the known weak donor power of the 2D material, with a consequently reduced back donation from the metal into CO ligands.

Another metal fragment for potential $\eta^2-P_n$ coordination is a $d^6-L_4M$ butterfly, obtained upon removal of two *cis* equatorial ligands from a $d^6-L_6M$ octahedron. A possible example is $(CO)_4M$ (M=Cr, Mo and W), obtainable for instance from an octahedron with two additional dialkyl sulphide ligands.[46,47] Optimization of the $(CO)_4Mo(\eta^2-P_n)$ adduct (see Figure S5) shows that the Mo-P interactions are relatively weak in view of the 2.56 Å distances and other evident distortions of steric origin. The scarce propensity of the butterfly $(CO)_4Mo$ fragment to support $P_n$ di-hapto coordination is corroborated by the large +1.02 eV endothermicity of $(CO)_4Mo(\eta^2-P_n)$, as estimated from Eq. 5. Perhaps in general, phosphorene functionalization is unlikely with a butterfly fragment.

$$(CO)_4Mo((CH_3)_2S)_2 + P_n \rightarrow (CO_4)Mo(\eta^2-P_n) + 2(CH_3)_2S \qquad \text{Eq. 5}$$

## $\eta^3$ Coordination Mode

The contiguous $P_3$ triangles at the $P_n$ surfaces (Figure 1) are attained by involving a single and a pair of P atoms belonging to different zig-zag chains. The triangles are isosceles but not dramatically far from being equilateral (3.62 and 3.35 Å for the two across-channel sides and the in-chain one). In principle, the three atoms in question represent an ensemble of *fac* ligands at a $d^6-L_6M$ octahedron, whose other three vertexes can be those of a $L_3M$ fragment, such as the $(CO)_3Mo$ one of the complex $(CO)_3Mo(\eta^6$-p-xylene),[48] whose poly-hapto aryl ring can be easily removed.

From the orbital interactions viewpoint, the unsaturated trigonal pyramidal (TP) $(CO)_3Mo$ fragment has the three vacant and delocalized MOs of $a_1 + e$ (degenerate) symmetries, shown in Scheme 3. These can match with phosphorene's lone pair combinations to allow $\eta^3-P_n$ coordination.

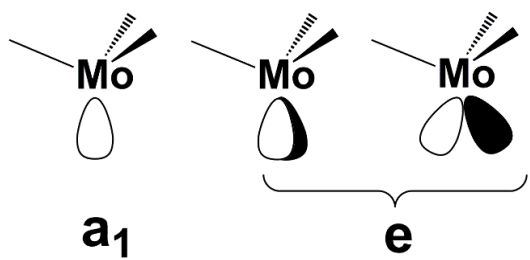

**Scheme 3.** Vacant FMOs of the uncharged $(CO)_3Mo$ fragment with their corresponding symmetries.

However, the e-symmetry matchings are not fully equivalent due to the natural isosceles shape of the $P_3$ triangle on the surface. In fact, the optimized $(CO)_3Mo(\eta^3\text{-}P_n)$ adduct of Figure 13 has the $P_1$-Mo bond of 2.46 Å, which is about 0.1 Å shorter than the $P_2$-Mo and $P_3$-Mo ones of 2.56 Å.

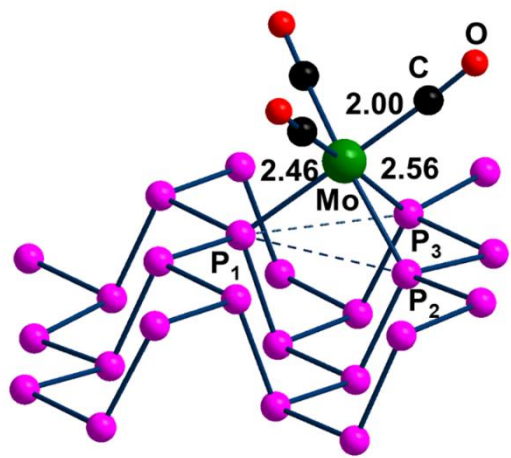

**Figure 13.** Optimized structure of the adduct $(CO)_3Mo(\eta^3\text{-}P_n)$.

As indicated by Scheme 4, the $P_1$ lone pair is properly oriented for optimal MO interactions with both $a_1$ and the e orbitals of Scheme 3. Conversely, the $P_2$ and $P_3$ lone pairs are not equally well oriented and can hardly rearrange due to the $P_n$ skeleton rigidity. Eventually, however, the divergences in bonding are not so dramatic, hence an essential tri-hapto coordination of phosphorene may be assumed.

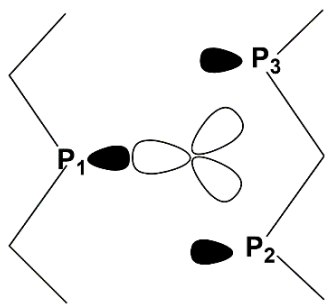

**Scheme 4.** Disposition of the vacant metal σ hybrids and the $P_1$, $P_2$ and $P_3$ lone pairs in a potential tri-hapto coordination.

From the energy viewpoint, the $\eta^3\text{-}P_n$ formation is exothermic by about -0.33 eV, as estimated by the reaction in Eq. 6, which proceeds from the precursor $(CO)_3Mo(\eta^3\text{-p-xylene})$.

$(CO)_3Mo(\eta^3\text{-p-xylene}) + P_n \rightarrow (CO)_3Mo(\eta^3\text{-}P_n) + \text{p-xylene}$  Eq. 6

Incidentally, the scarce deformation of the coordinated $P_n$ vs. the free 2D material has a cost of +0.28 eV, as it results from a single point calculation of the final adduct upon removal of the $(CO)_3Mo$ fragments. Evidently, the formed P-Mo bonds almost double the energy lost on the $P_n$ deformation. In addition, the discrete complex $(CO)_3Mo(PMe_3)_3$ was optimized to compare its computed IR spectrum with that of $(CO)_3Mo(\eta^3-P_n)$. For consistency, the vibrational frequencies of all species were calculated by using the same CRYSTAL package.[19] Once again, the donor power of phosphorene is found to be scarce as indicated by the 60 cm$^{-1}$ blue shift of the CO stretching in the adduct vs. those of the complex $(CO)_3Mo(PMe_3)_3$. Consider that in latter the carbo-substituted phosphines have high donor power, so that the larger electron density at the metal allows larger back-donation with different IR response for CO stretching. As another point, adducts of the type $L_3Mo(\eta^3-P_n)$ with three single bulky substituted phosphines[49] or the common tripodal chelate such as $CH_3C(CH_2)_3PPh_2$[50] are unlikely because of the steric hindrance problems arising when the substituents point down on the $P_n$ surface.

Finally, as already done for the $P_n$ $\eta^1$ coordination with the ClAu fragment, we examined how the electronic structure and the determining band gap may be influenced by the $(CO)_3Mo$ functionalization. The corresponding gap of 2.34 eV for $(CO)_3Mo(\eta^3-P_n)$ is larger than in free phosphorene and the discussed $ClAu(\eta^1-P_n)$ adduct (2.26 and 2.17 eV, respectively). At variance with the latter case, the P lone pairs, which participate in the P-Mo bonding, lose their reciprocal repulsive character, with a lowering of the valence band spread, and consequent increasing of the band gap. On the other hand, the right side of Figure S6 shows that in both the valence and conduction bands, the π interactions between the $t_{2g}$ metal d orbitals and CO π* levels have significant contribution. In particular, the stabilized Mo d orbitals represent the highest limit of the valence band, while the opposite occurs for $ClAu(\eta^1-P_n)$, where the repulsions between the low lying and filled Au d orbitals and the P lone pairs fix the band gap.

**Conclusions**

The article has addressed potential cases of the phosphorene's covalent functionalization, stemming from the compactness of the P lone pairs at the surface. Their behavior as single or combined 2e$^-$ donors has been underlined concerning the interactions with acidic main group units or unsaturated transition metal fragments. Interesting electronic aspects have emerged from the solid state calculations compared with related molecular models. All the interactions are of the acid-base Lewis type, even though the P lone pairs of $P_n$ surface feature limited donor power. A possible reason for this is that the lone pairs participating in the valence bands are not separated but somewhat mixed into each other throughout the surface. Accordingly, such a delocalization does not allow maximized σ overlap with the added acidic groups. Notice, that in this case the invoked delocalization is very different from the π one in planar graphene, with determining π bonding to be excluded for phosphorene. These points have clearly emerged for instance from the adducts with the acidic $BH_3$ molecule and, in particular, those having variable coverage of the surface. Di-iodine has been also tested to check whether it can induce P-P cleavage at phosphorene, as we recently found for white phosphorus ($P_4$).[12] This hypothesis had to be excluded, because $I_2$ (or two combined molecules of it) in no case has the possibility of attacking from outside any P-P σ* level with consequent P-P cleavage.

Next, we switched to various transition metal fragments featuring a variable number of vacant metal lobes, based on the *isolobal analogy* concept.[30] Accordingly, we examined the possibility that some given metal acceptor interacts with one or more neighbor $P_n$ lone pairs and afford its covalent functionalization. M-$P_n$ derivatives of various hapticity were optimized providing useful information on stereochemistry and energy. Evident steric hindrance problems emerge even when a metal fragment is electronically suited but has an unsuitable disposition of the coligands, in particular if bulky. Other derived properties of the adducts have been examined such as the vibrational spectra of the CO coligands, when present. Another tackled point concerns the band gap of the functionalized adduct *vs*. that of naked phosphorene as a consequence of the variable repulsion between P lone pairs. A first example of perturbation is given by the stacking of the $P_n$ sheets to reform black phosphorus. Only ten $P_n$ sheets are sufficient to reproduce the much smaller band gap of the 3D material, because the bilateral repulsions between the sheets quickly spread frontier band and reduce its separation from the conduction band. On the other hand, the functionalization of a single $P_n$ face has contrasting and discussed effects such as the widening of the ditches on associating acidic groups, which in turn reduces the repulsion between lone pairs. In cases of metal coordination, its d orbital set may fall at the top of the bands formed by P lone pairs with consequent modification of the ultimate band gap. Qualitative explanations have been proposed also for more complex cases.

While our analysis has indicated various possibilities of covalent functionalization of phosphorene with acidic groups, the corresponding experimental evidence remains scarce. Certainly, one reason is the difficult isolation and manipulation of the 2D material, given its known sensitivity to external agents such as oxygen or humidity. A preliminary elimination of the latter is fundamental to perform new useful chemistry. In any case, we trust that our models may be later exploited by some brave experimentalist to construct viable species and open new application gates in the chemistry of the important 2D material.

**Computational Details**

The optimized geometries and energetics of the naked and functionalized phosphorene surface have been established at B3LYP-DFT[21] level of theory by using the CRYSTAL17 package.[19] A selected supercell of 16 phosphorus atoms has been usually adopted in performing the optimization of both the atomic positions and lattice parameters. The TZVP basis set[51] has been used for all the atomic species. Only for the elements: Ru,[52] Mo,[53] for Au[54] and I[55] different *pseudo-potential* have been employed. In the metal adducts with CO ligands, the vibrational frequencies have been carried out to compare the donor power of the $P_n$ surface with those of other P donors. Band and Crystal Overlap Orbital Populations (COOP) analysis have been carried out with the available routines of the CRYSTAL17 package.[19] List of all the optimized structural and energy parameters are available in the Supporting Information.

**ACKNOWLEDGMENTS**

The authors thank the European Research Council (ERC) under the European Union's Horizon 2020 research and innovation program (Grant Agreement No. 670173) for funding the project **PHOS*FUN* "*Phos*phorene *fun*ctionalization: a new platform for advanced multifunctional materials*" through an ERC Advanced Grant. Thanks are due also to ISCRA-CINECA HP Grants HP10C2QI78 and HP10CFMSCC for the computational resources.